\documentclass[pla,reprint,onecolumn]{elsarticle}
\journal{Indian Journal of Physics}
\usepackage{amsmath,amssymb,physics,braket,subfloat,graphicx,subfigure,epstopdf,lastpage,hyperref}
\hypersetup{colorlinks=true, urlcolor=red, citecolor=blue, linkcolor=blue}
\allowdisplaybreaks
\begin{document}
\title{A comparative study of higher-order nonclassicalities of photon-added-then-subtracted and photon-subtracted-then-added quantum states}
\author{Deepak}
\ead{deepak20dalal19@gmail.com}

\author{Arpita Chatterjee\corref{cor1}}
\ead{arpita.sps@gmail.com}

\cortext[cor1]{Corresponding author}
\date{\today}
\address{Department of Mathematics, J. C. Bose University of Science and Technology,\\ YMCA, Faridabad 121006, India}
\begin{abstract}
In the present paper, we have studied the higher as well as the lower-order nonclassicalities of photon-added-then-subtracted and photon-subtracted-then-added thermal and even coherent states. Different criteria such as Mandel's function ($Q_M^{(l)}$), higher-order antibunching ($d_h^{(l-1)}$), sub-Poissonian photon statistics ($D_h^{(l-1)}$), higher-order squeezing ($S^{(l)}$), Husimi function ($Q$), Agarwal-Tara criteria ($A_3$) and Klyshko's condition ($B(m)$) are used to witness the nonclassical feature of these states. Many of these conditions established that the considered states are highly nonclassical. It is also realized that the non-Gaussian photon-addition-then-subtraction operation is preferred over the photon-subtraction-then-addition for developing nonclassicality.
\end{abstract}
\maketitle

\section{Introduction}
\label{intr}
Quantum physics has long been fascinated by the non-commutativity of the annihilation $(a)$ and creation $(a^\dagger)$ operators \cite{whl}. Simple alternated sequences of adding and subtracting identical photons from any quantum system produce different results due to the non-commutativity of bosonic operators. Recently, Zavatta et al. \cite{jmo11,jmo12,jmo13} have proposed a successful experiment to control photon subtraction from or addition to a light beam using straightforward optical processes like beam splitting, frequency down-conversion, and homodyne detection. The design of nonclassical states with various practical applications is now possible due to these experimental successes. For instance, entangled states are used to realize quantum computing and to transmit quantum information \cite{jmo15}, squeezed states are used to reduce the noise level in one of the phase-space quadratures below the quantum limit \cite{jmo14}, etc. It has been experimentally demonstrated that in addition to facilitating quantum communication, subtracting a photon from one of the two modes of a two-mode squeezed state can enhance entanglement \cite{jmo9}. 
Many significant applications in quantum information science are based on the manipulation of a light field at the single-photon level. In the framework of a nonclassicality test \cite{ln2} or an entanglement distillation \cite{ln3}, two elementary operations on a single-mode field (i.e., photon subtraction and addition represented by bosonic annihilation $a$ and creation $a^\dagger$ operators, respectively) can be applied to specifically change a field state into a desired one. For instance, photon subtraction converts a Gaussian entangled state (two-mode squeezed state) to a non-Gaussian one. Both the photon-subtracted
\cite{ln5,ln6,ln7,ln8} and photon-added squeezed states \cite{ln9,ln10} are proposed to improve the fidelity of continuous variable (CV) teleportation. The photon addition is known to create a nonclassical state from any classical state (e.g., coherent and thermal states) \cite{rep8.1}. The addition \cite{ln11} and subtraction \cite{jmo11} of photons are now practically realized in laboratories. A simple sequence of adding and subtracting identical particles to any quantum system produces different outputs if the order is interchanged. By using the bosonic operations, multiple numbers of photons can be added (subtracted) to (from) the quantum state under consideration \cite{priyam,pmalpani}. This technique of photon addition by using the boson operators was introduced by Agarwal and Tara \cite{rep8.1}.
A beam splitter can also be used to make a photon-added or photon-subtracted state \cite{rep8.2,ren}. Dakna et al. \cite{rep8.3} proved that if a number state and any initial quantum state are injected into two input channels of a beam-splitter, the counting of photons in one output mode decreases (increases) and in second output mode, the corresponding number of photons increases (decreases), caused the resultant photon-added (subtracted) output state.
Recently Parigi et al. \cite{rep8.5} successfully constructed an experimental set-up for subtracting (adding) a single photon from (to) a thermal light field that is completely classical and incoherent. By repeating the above procedure, they further developed the method for subtracting (adding) multiple photons from (to) any state.
These experimental visualizations motivated us to investigate the photon addition and subtraction phenomena in detail. Moreover, the degaussification process can be acquired when photons are added (subtracted) to (from) a classical Gaussian radiation field, and then the resulting states display nonclassicality \cite{usha}. Yang and Li \cite{yang} further reported that the nonclassicality of the initial Gaussian state can be enhanced by the photon subtraction process. They emphasized that the states generated by first adding (subtracting) multi-photons to an arbitrary state (whether classical or nonclassical) and then subtracting (adding) multi-photons from the resulting state is certainly nonclassical if the number of added photons is equal to or greater than the number of subtracted photons. Motivated by these facts, we consider two important Gaussian field states and study their nonclassical features introduced by adding and subtracting a specific number of photons. We work on different possibilities such as photon addition number = ($>$ or $<$) photon subtraction number. It is important to note that the nonclassicalities of such engineered states have been studied mostly by using a set of lower-order witnesses. Due to this, we aim to check how the changes in photon addition and subtraction numbers affect the higher-order nonclassicality parameters.

The coherent state \cite{cs} is considered as the most classical-like quantum state because its Glauber-Sudarshan $P$ function is exactly equal to $\frac{1}{2}$. For this unique nature, the coherent state has many applications in different fields of Science \cite{ppr1}. Mathematically, the coherent state is defined as the eigenstate of bosonic annihilation operator $a$ \cite{whl}, i.e., if $\ket{\gamma}$ denotes the usual coherent state then $a\ket\gamma=\gamma\ket\gamma$. The coherent state can also be generated by applying the displacement operator on Fock state $\ket{0}$ \cite{book}. {Quantum oscillators prepared out of thermal equilibrium can be used to produce work and transmit information. Thermal state of a cooled system is not in thermal equilibrium with its environment and can be used to perform work and carry information. In quantum mechanics, even coherent states have several potential benefits over odd coherent states, such as a larger overlap with the ground state, a higher probability density near the origin, a simpler mathematical form, and easier generation using certain experimental techniques. Thus we have chosen these two particular states for further investigation.} The density matrix $\sigma^{\mathrm{th}}$ of thermal state \cite{dm, ts} is expressed in terms of Fock states as follows $\sigma^{\mathrm{th}}=\frac{1}{(1+\bar{n})}\sum_{n=0}^\infty\left(\frac{\bar{n}}{1+\bar{n}}\right)^n\ket{n}\bra{n}$, where $\bar{n}$ is the mean photon number. The even coherent state \cite{ecs} can be defined by the superposition of two coherent states of equal magnitude but opposite direction, i.e. $\ket\psi^{ecs} = \frac{1}{(1+2e^{-|\alpha|^2})}(\ket\alpha+\ket{-\alpha})$. Here, we focus on the higher-order nonclassical properties of engineered thermal and even coherent states.

{ Higher-order nonclassicality \cite{alam2} is a concept in quantum mechanics that refers to the degree of nonclassical character exhibited by a quantum state. It measures the deviation of a quantum state from classical probability distributions and reflects the extent to which the state violates classical notions of reality. Higher-order nonclassicality can be used to characterize the unique features of quantum systems that distinguish them from classical systems. For example, higher-order nonclassicality can be used to quantify the degree of entanglement between two or more particles, which is a crucial resource in quantum information processing tasks such as quantum teleportation and quantum cryptography. Moreover, higher-order nonclassicality can be used to describe the properties of exotic quantum states, such as squeezed states, cat states, and coherent states, which have no classical analogues. These states have many applications in different fields, including quantum metrology, quantum computation, and quantum communication. Higher-order nonclassicality is also relevant for understanding the fundamental principles of quantum mechanics including the measurement problem and the interpretation of quantum states. It is noted that highly nonclassical states can manifest unusual behavior that challenges our classical intuitions about the nature of reality, leading to a deeper understanding of the underlying principles of quantum mechanics.}

In recent times, Priya et al. \cite{priyam,pmalpani} used the criteria of higher-order nonclassicality to detect the nonclassical nature of photon-added and subtracted displaced Fock states. Further, these higher-order criteria are used to detect the nonclassicality for a coherent superposed quantum state \cite{ppr1} and SUP-engineered coherent and thermal states \cite{ppr2}. It is observed that higher-order criteria are more useful to detect the weaker nonclassicality of different engineered quantum states. The same group of people have studied the nonclassical properties of photon-added-then-subtracted displaced Fock state in \cite{pngp}. 
{ Photon-added-then-subtracted (PAS) and photon-subtracted-then-added (PSA) thermal states and even coherent states are quantum states that have applications in various fields of quantum physics \cite{r32,r33}. In quantum optics, PAS and PSA thermal states are used to describe the photon number statistics of a system interacting with a thermal environment. These states have been employed to study the quantum-to-classical transition in the dynamics of open quantum systems as well as the properties of squeezed states of light \cite{r34,r35}. Similarly, even coherent states are widely used in the study of quantum phase transitions and the analysis of Bose-Einstein condensates. In quantum information theory, these states are also applied in quantum communication and cryptography. Specifically, PAS and PSA thermal states have been used to study the security of quantum key distribution protocols, while even coherent states have been employed in the design of continuous-variable quantum cryptographic protocols. Overall, these quantum states are proved to be useful for exploring the fundamental principles of quantum mechanics as well as for the development of practical applications in quantum information processing and quantum communication.} Thus we will study the nonclassical properties of photon-added-then-subtracted and photon-subtracted-then-added thermal and even coherent states. Further, we will compare the changes in nonclassical behavior of quantum states by adding $p$ photons then subtracting $q$ photons, and operating the same operation in reverse order i.e. by subtracting $q$ photons then adding $p$ photons. So in the present work, we aim to use the higher-order criteria for identifying the nonclassicality of photon-added-then-subtracted (PAS) and photon-subtracted-then-added (PSA) thermal and even coherent states. The PAS (PSA) operation is defined by adding (subtracting) $q$ ($p$) photons to (from) the quantum state and then subtracting (adding) $p$ ($q$) photons from (to) the resulting state, that means $A=a^pa^{\dagger q}$ ($B =a^{\dagger q}a^p$). The nonclassicality of photon-added-then-subtracted (photon-subtracted-then-added) thermal states (PAST (PSAT)) and even coherent states (PASEC (PSAEC)) are discussed in \cite{rep8} by using the photon number distribution and Wigner function. Here we extend that work by considering the higher-order measures \cite{priyam,ppr1,ppr2} namely higher-order Mandel's function \cite{ho51}, higher-order anti-bunching (HOA) \cite{ho29}, higher-order sub-poissonian photon statistics (HOSPS) \cite{ho30}, higher-order squeezing (HOS) \cite{ho31}, Husimi $Q$ function \cite{ho48}, Agarwal-Tara criteria \cite{pmalpani} and Klyshko's condition \cite{ho66}. We perform a comparative study of the nonclassical features of PAS and PSA-operated thermal and even coherent states.

The present paper is structured in the following manner: we study the PAS (PSA)-operated thermal and even coherent states in Sec.~\ref{qsi}. We compare the nonclassicality of PAS and PSA-operated thermal and even coherent states in Sec.~\ref{ncc}. The next section ends with a conclusion.

\section{Quantum states of interest}
\label{qsi}
	
The density matrix for the well-known thermal state is given by \cite{rep8}
	\begin{equation}
		\label{th}
		\sigma^{ts} = \frac{1}{1+\bar{r}}\sum_{r=0}^\infty\left(\frac{\bar{r}}{1+\bar{r}}\right)^r\ket{r}\bra{r},
	\end{equation}
	$\bar{r}$ is the mean photon number. Applying the operators $A=a^pa^{\dagger q}$ and $B=a^{\dagger q}a^p$ on the thermal state, we obtain the density matrices of photon-added-then-subtracted (PAST) and photon-subtracted-then-added (PSAT) thermal states, respectively
	\begin{equation}
		\label{past}
		\sigma_1^{ts}=A\sigma^{th}A^\dagger=\frac{N_1^{ts}}{1+\bar{r}}\sum_{r=0}^{\infty}\left(\frac{\bar{r}}{1+\bar{r}}\right)^r\frac{(r+q)!^2}{r!(r+q-p)!}\ket{r+q-p}\bra{r+q-p},
	\end{equation}
	and
	\begin{equation}
		\label{psat}
		\sigma_2^{ts}=B\sigma^{th}B^\dagger=\frac{N_2^{ts}}{1+\bar{r}}\sum_{r=0}^{\infty}\left(\frac{\bar{r}}{1+\bar{r}}\right)^r\frac{r!(r-p+q)!}{(r-p)!^2}\ket{r-p+q}\bra{r-p+q},
	\end{equation}
	
where the normalization constants $N_1^{ts}$ and $N_2^{ts}$ are given by \cite{rep8} $$(N_1^{ts})^{-1} = \left\{
	\begin{array}{ll}
		\frac{q!^2}{(1+\bar{r})(q-p)!}~_2F_1\left(1+q,1+q;1+q-p;\frac{\bar{r}}{1+\bar{r}}\right) & \mbox{if } q\geq p, \\\\
		\frac{p!^2}{(1+\bar{r})(p-q)!}~_2F_1\left(1+p,1+p;1+p-q;\frac{\bar{r}}{1+\bar{r}}\right) &\mbox{if } q<p
	\end{array}
	\right.$$ and $$(N_2^{ts})^{-1}=\frac{p!q!}{1+\bar{r}}\left(\frac{\bar{r}}{1+\bar{r}}\right)^p~_2F_1\left(1+q,1+p;1;\frac{\bar{r}}{1+\bar{r}}\right),$$
	$ _pF_q(a_i;b_j;c)$ is the well-known generalized hypergeometric function.
	
By applying $A$ ($B$) on the even coherent state $\ket\psi^{ecs}=\frac{1}{(2+2e^{-2|\alpha|^2})^{1/2}}(\ket{\alpha}+\ket{-\alpha})$, we get $\ket\psi_1^{ecs}$ ($\ket\psi_2^{ecs}$) states as follows \cite{rep8}
	\begin{equation}
		\ket\psi_1^{ecs}={N_1^{ecs}a^pa^{\dagger q}}(\ket\alpha+\ket{-\alpha}),
	\end{equation}
	and
	\begin{equation}
		\ket\psi_2^{ecs}={N_2^{ecs}a^{\dagger q}a^p}(\ket\alpha+\ket{-\alpha}),
	\end{equation}
$N_1^{ecs}$ and $N_2^{ecs}$ are the corresponding normalization constants \cite{rep8}.

\section{Nonclassicality criteria}
\label{ncc}
	
	The expectations of $\langle a^{\dagger m}a^n\rangle$ with respect to the states $\sigma_1^{ts}$, $\sigma_2^{ts}$, $\ket{\psi}_1^{ecs}$ and $\ket{\psi}_2^{ecs}$ are as follows (see \ref{a1} for detailed calculation).

	\begin{eqnarray}
		E_1^{ts} = \left\{
		\begin{array}{lll}
			\frac{N_1^{ts}q!^2}{(q-p-n)!(1+\bar{r})}\\\times~_2F_1\left(1+q,1+q;q-p-n+1;\frac{\bar{r}}{1+\bar{r}}\right)

			\delta_{m,n} & \mbox{if} & q-p-n\geq 0\\\\
			\frac{N_1^{ts}(p+n)!^2}{(p-q+n)!(1+\bar{r})}\left(\frac{\bar{r}}{1+\bar{r}}\right)^{p-q+n}\\\times~_2F_1\left(1+p+n,1+p+n;p-q+n+1;\frac{\bar{r}}{1+\bar{r}}\right)

			\delta_{m,n} & \mbox{if} & q-p-n < 0
		\end{array}
		\right.
		\label{exppast}
	\end{eqnarray}
	\begin{equation}
		E_2^{ts} = \left\{
		\begin{array}{lll}
			\frac{N_2^{ts}}{1+\bar{r}}\left(\frac{\bar{r}}{1+\bar{r}}\right)^p\frac{q!^2p!}{(q-n)!}~_3F_2\left(1+q,1+q,1+p;1,q-n+1;\frac{\bar{r}}{1+\bar{r}}\right)
			\delta_{m,n} &\mbox{if } q\geq n \\\\			\frac{N_2^{ts}}{1+\bar{r}}\left(\frac{\bar{r}}{1+\bar{r}}\right)^{p-q+n}\frac{n!^2(p-q+n)!}{(n-q)!^2}\\
\times~_3F_2\left(1+n,1+n,1+p-q+n;1-q+n,1-q+n;\frac{\bar{r}}{1+\bar{r}}\right)
			\delta_{m,n} &\mbox{if } q < n
		\end{array}
		\right.
		\label{exppsat}
	\end{equation}
	$E_1^{ts}$ and $E_2^{ts}$ are the expectations $\langle a^{\dagger m}a^n\rangle$ with respect to $\sigma_1^{ts}$ and $\sigma_2^{ts}$, respectively, $\delta_{mn}$ is the Kronecker delta defined as \cite{kdel}
	\begin{equation}
	\braket{m|n}=\delta_{mn}=\left\{
	\begin{array}{ll}
	1 &\mbox{ if } m=n\\\\
	0 &\mbox{ elsewhere }
	\end{array}\right..
	\end{equation}
	Again,
	\begin{align}
		\label{eq:exppasec}
		\langle a^{\dagger m}a^n\rangle_1^{ecs}& = (N_1^{ecs})^2\sum_{r=0}^{\text{min}(n+p,q)}{n+p\choose r}{q\choose r}r!\alpha^{n+p-r}(-1)^q\Big[H_{m+p+q-r,q}(\alpha^*,-\alpha)\\&+e^{-2|\alpha|^2}H_{m+p+q-r,q}(-\alpha^*,-\alpha)\Big]\nonumber
	\end{align}
	where $H_{m,n}(x,y)$ is the usual bivariate Hermite polynomial \cite{rep8} and
	\begin{align}
		\label{eq:exppsaec}
		\langle a^{\dagger m}a^n\rangle_2^{ecs}&=(N_2^{ecs})^2|\alpha|^{2q}\sum_{r=0}^{\text{min}(n,q)}{n\choose r}{q\choose r}r!\alpha^{n-r}(-1)^q\Big[H_{m+q-r,q}(\alpha^*,-\alpha)\\&+e^{-2|\alpha|^2}H_{m+q-r,q}(-\alpha^*,-\alpha)\Big]\nonumber
	\end{align}
	\subsection{Higher-order Mandel's function}
	
\begin{figure}[htb]
		\centering
		\includegraphics[scale=.9]{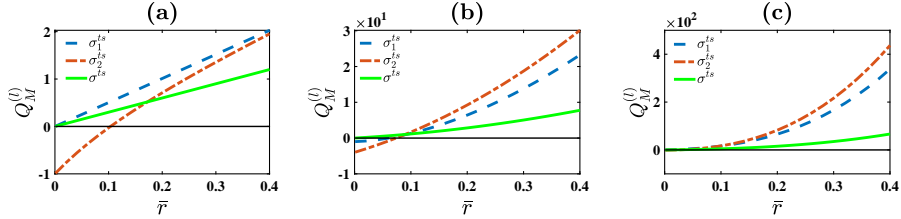}
		\caption{(Color online) $Q_M^{(l)}$ with respect to mean photon number $\bar{r}$ for (a) $l=2,\,p=1,\,q=1$, (b) $l=3,\,p=1,\,q=2$ and (c) $l=4,\,p=2,\,q=1$. The green solid line corresponds to the initial thermal state $\sigma^{ts}$ ($p=q=0$) with respective $l$ values.}
		\label{qmltfig}
	\end{figure}

\begin{figure}[htb]
		\centering
		\includegraphics[scale=.9]{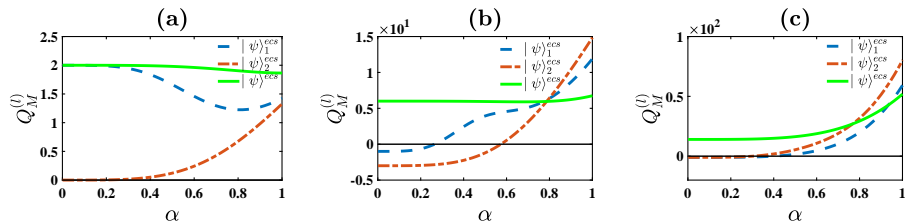}
		\caption{(Color online) $Q_M^{(l)}$ with respect to even coherent state parameter $\alpha$ for (a) $l=2,\,p=1,\,q=1$, (b) $l=3,\,p=1,\,q=2$ and (c) $l=4,\,p=2,\,q=1$. The green solid line corresponds to the initial even coherent state $\ket{\psi}^{ecs}$ ($p=q=0$) with respective $l$ values.}
		\label{qmlecfig}
	\end{figure}

One of the criteria that is used to detect the nonclassicality of any arbitrary quantum state is known as Mandel's $Q_M$ parameter or function \cite{mandel}. This can be defined in terms of the photon number distribution up to arbitrary order $l$ as follows \cite{sanjib}
	
	\begin{eqnarray}
		\label{qml}
		Q_M^{(l)} & = & \frac{\langle{(\Delta{\mathcal{N}})^l}\rangle}{\langle{a^\dagger a}\rangle}-1,
	\end{eqnarray}
	where $\Delta{\mathcal{N}}\,=\,a^\dagger a-\langle{a^\dagger a}\rangle$ is the dispersion in the number operator $\mathcal{N}=a^\dagger a$. Using the identity \cite{sanjib}
	\begin{eqnarray*}
		\langle{(\Delta{\mathcal{N}})^l}\rangle = \sum_{k=0}^l {l \choose k}(-1)^k\langle (a^\dagger a)^{l-k}\rangle{\langle a^\dagger a\rangle}^k
	\end{eqnarray*}
	and \cite{moya1}
	\begin{equation}\nonumber
		(a^\dagger a)^r = \sum_{n = 0}^r  S_r^{(n)}a^{\dagger n}a^n,
	\end{equation}
	where $S_r^{(n)}$ is the Stirling number of second kind \cite{stegun}
	\begin{equation}\nonumber
		S_r^{(n)} = \frac{1}{n!}\sum_{j=0}^n (-1)^{n-j}{n\choose j}j^r,
	\end{equation}
	The Mandel's parameter $Q_M^{(l)}$ can be evaluated explicitly up to order $l$.  All expectations in \eqref{qml} are calculated using \eqref{exppast}, \eqref{exppsat}, \eqref{eq:exppasec} and \eqref{eq:exppsaec} for $\sigma_1^{ts}$, $\sigma_2^{ts}$, $\ket\psi_1^{ecs}$ and $\ket\psi_2^{ecs}$ states, respectively. The parameter $Q_M^{(l)}$ is plotted with respect to the mean photon number $\bar{r}$ in Fig.~\ref{qmltfig} for thermal state and the coherent state amplitude $\alpha$ in Fig.~\ref{qmlecfig} for even coherent state.
	
In Fig.~\ref{qmltfig}, it can be easily seen that the maximum depth of nonclassicality is observed by the lower-order $Q_M^{(l)}$ ($l=2$) parameter for the PSAT state whereas nonclasscality of PAST could not be detected by it when an equal number of photons are added and subtracted (see Fig.~\ref{qmltfig}(a)). Further, the higher-order criteria ($l=3$) detects the nonclassicality of both the PAST and PSAT states only when the number of photons subtracted is less than the number of photons added and in this case, the depth of nonclassicality for the PSAT state is more than PAST state. In case the number of photons added is less than the photons subtracted, then the higher-order nonclassicality criteria ($l=4$) fails to detect the nonclassicality of both PAST and PSAT states. Also, the depth of nonclassicality decreases with increasing mean photon number for all lower as well as higher-order criteria. Hence, we can conclude that Mandel's $Q$ parameter detects the nonclassicality of PAST and PSAT states when the addition number of photons is greater than or equal to the subtraction number which is in accordance with Yang et al. \cite{yang} and shows that the depth of nonclassicality for the PAST state is less than the PSAT state. { The nonclassicality of the original thermal state is not detected by the lower as well as higher-order Mandel's $Q$ function.} Similar observations can be drawn for even coherent states (see Fig.~\ref{qmlecfig}). Here the lower-order criteria ($l=2$) fails to detect the nonclassicality of both PASEC and PSAEC for $p=q=1$. But the higher-order witness ($l=3$, $4$) detects the nonclassicality of both PASEC and PSAEC states when the number of photons added is greater than the number of photons subtracted and shows that the depth of nonclassicality for PSAEC is more than PASEC (see Figs.~\ref{qmlecfig}(b)-(c)). Also the nonclassicality of both the states decreases with increasing even coherent state parameter $\alpha$. { Further, the nonclassicality of the original even coherent state is detected neither by lower nor by higher-order Mandel's` function.}

	\subsection{Higher-order anti-bunching}
	
\begin{figure}[htb]
		\centering
		\includegraphics[scale=.9]{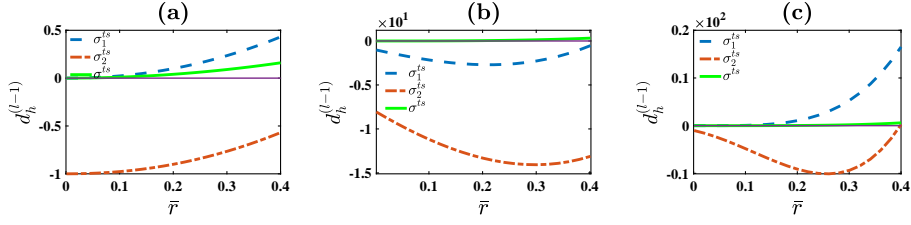}
		\caption{(Color online) $d_h^{(l-1)}$ with respect to thermal state mean photon number $\bar{r}$ for (a) $l=2,\,p=1,\,q=1$, (b) $l=3,\,p=1,\,q=2$ and (c) $l=4,\,p=2,\,q=1$.}
		\label{dntfig}
	\end{figure}

\begin{figure}[htb]
		\centering
		\includegraphics[scale=.9]{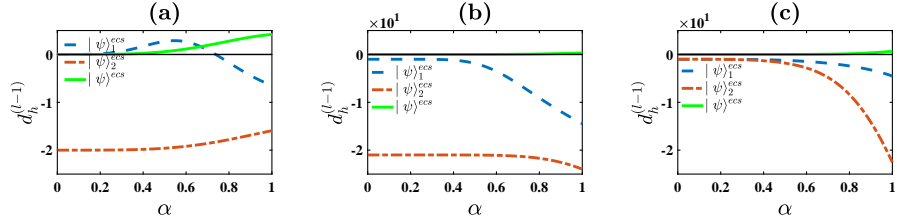}
		\caption{(Color online) $d_h^{(l-1)}$ with respect to even coherent state amplitude $\alpha$ for (a) $l=2,\,p=1,\,q=1$, (b) $l=3,\,p=1,\,q=2$ and (c) $l=4,\,p=2,\,q=1$.}
		\label{dnecfig}
	\end{figure}

	Initially, the concept of HOA was introduced by Lee \cite{lee} to detect the nonclassicality of any quantum state. Pathak and Gracia \citep{ho29} further modified the concept of HOA for improving its physical meaning. The $(l-1)$-th order HOA is given by

	\begin{equation}
		\label{dn}
		d_h^{(l-1)} = \braket{a^{\dagger l}a^l}-\braket{a^\dagger a}^l~<~0,
	\end{equation}
that means the probability of coming photons bunched is less than photons coming independently. It is an excellent candidate for identifying the nonclassicality of any quantum state. The expectations in \eqref{dn} can be calculated by using \eqref{exppast}, \eqref{exppsat}, \eqref{eq:exppasec} and \eqref{eq:exppsaec} for $\sigma_1^{ts}$, $\sigma_2^{ts}$, $\ket\psi_1^{ecs}$ and $\ket\psi_2^{ecs}$ states respectively. The corresponding graphs are plotted in Fig.~\ref{dntfig} and Fig.~\ref{dnecfig}.
	
Figure \ref{dntfig} shows the dependence of anti-bunching parameter $d_h^{(l-1)}$ on mean photon number $\bar{r}$ of PAST and PSAT states for different addition and subtraction photon numbers. The lower-order $d_h^{(l-1)}$ shows a highly deep region for PSAT while a partially deep curve for PAST when $p=q=1$. The depth of nonclassicality decrease for both states as $l$ changes from 3 to 4. From Figs.~\ref{dntfig} and \ref{dnecfig}, it is clear that $\sigma_2^{ts}$ $(\ket\psi_2^{ecs})$ is more nonclassical than $\sigma_1^{ts}$ $(\ket\psi_1^{ecs})$ as the negativity of $d_h^{(l-1)}$ for the first one is more as compared to the later. Also for any number of photon addition or subtraction, the region of $\bar{r}$ ($\alpha$) is greater, when the operator $B$ is applied to a thermal (even coherent) state. Again from both the Figs.~\ref{dntfig} and \ref{dnecfig}, $d_h^{(l-1)}$ is always less for PSA-operated as compared to PAS-operated thermal (even coherent) state for arbitrary values of $\bar{r}$ ($\alpha$). In the case of PAST or PSAT, the depth of nonclassicality first increases with increasing mean photon number as $l$ changes from 2 to 3, and after a fixed value of $\bar{r}$, it starts decreasing. Further the depth of nonclassicality for both PSAEC and PASEC states, detected by the higher-order $d_h^{(l-1)}$, increases with increasing $\alpha$. { Similar to the Mandel's $Q$ function, HOA merely detects the nonclassicality of original thermal and even coherent states.}

\subsection{Higher-order sub-Poissonian photon statistics}
	
\begin{figure}[htb]
		\centering
		\includegraphics[scale=.9]{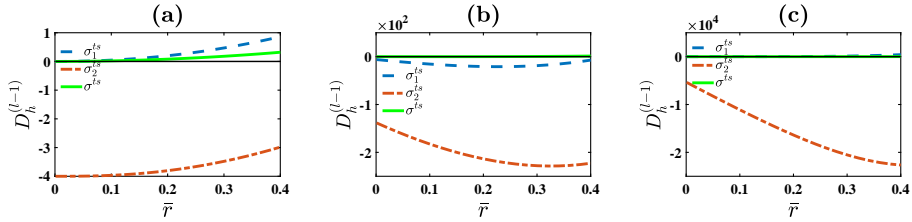}
		\caption{(Color online) $\mathcal{D}_h^{(l-1)}$ with respect to mean photon number $\bar{r}$ for (a) $l=2,\,p=1,\,q=1$, (b) $l=3,\,p=1,\,q=2$ and (c) $l=4,\,p=2,\,q=1$.}
		\label{dhltfig}
	\end{figure}

\begin{figure}[htb]
		\centering
		\includegraphics[scale=.9]{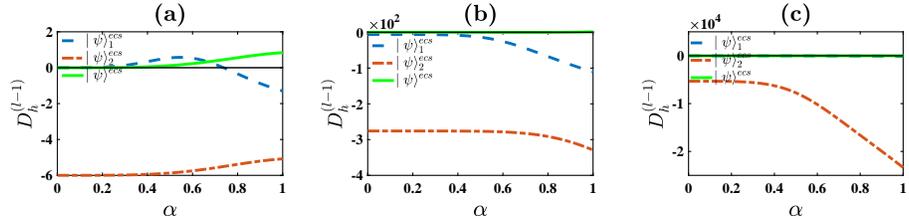}
		\caption{(Color online) $\mathcal{D}_h^{(l-1)}$ with respect to even coherent state amplitude $\alpha$ for (a) $l=2,\,p=1,\,q=1$, (b) $l=3,\,p=1,\,q=2$ and (c) $l=4,\,p=2,\,q=1$.}
		\label{dhlecfig}
	\end{figure}

{ The existence of higher-order nonclassicality in radiation fields is confirmed by the crucial feature of HOSPS. Although the lower-order phenomena of antibunching and sub-Poissonian photon statistics are closely linked, recent developments in this direction have shown that they can occur independently of each other \cite{kishore1,kishore3}. This means that the presence of sub-Poissonian photon statistics does not necessarily imply the occurrence of anti-bunching, and vice versa. Moreover, the higher-order nonclassicality and sub-Poissonian photon statistics can exist even if their lower-order counterparts are absent \cite{alam1} which highlights the need for careful analysis of HOA as well as HOSPS.} The generalized criteria for observing the $(l-1)$-th order sub-Poissonian photon statistics (for which $\langle(\Delta\mathcal{N})^l\rangle < \langle(\Delta\mathcal{N})^l\rangle_{\ket{\mathrm{Poissonian}}}$) is given by \cite{amit}
	
	\begin{equation}
		\mathcal{D}_h^{(l-1)} = \sum_{e=0}^{l}\sum_{f=1}^{e}S_2(e,f)^lC_e(-1)^ed_h^{(f-1)}{\langle a^\dagger a \rangle }^{l-e}\,\,<\,\,0
		\label{dhl}
	\end{equation}
	where $S_2(e, f)=\sum_{r=0}^{f} {^fC_r}(-1)^r r^e$ is the Stirling number of second kind, $^lC_e$ is the usual binomial coefficient.
	The analytic expressions of HOSPS for the states $\sigma_1^{ts}$, $\sigma_2^{ts}$, $\ket\psi_1^{ecs}$ and $\ket\psi_2^{ecs}$ can be obtained by substituting \eqref{past}, \eqref{exppsat}, \eqref{eq:exppasec} and \eqref{eq:exppsaec} in \eqref{dhl} and the simplified expressions are plotted in Figs.~\ref{dhltfig} and \ref{dhlecfig}.
	
	
	From the Figs.~\ref{dhltfig} and \ref{dhlecfig}, it is clear that $\sigma_2^{ts}$ ($\ket\psi_2^{ecs}$) shows higher nonclassicality than $\sigma_1^{ts}$ ($\ket\psi_1^{ecs}$), because $\mathcal{D}_h^{(l-1)}$ for $\sigma_2^{ts}$ ($\ket\psi_2^{ecs}$) is much more negative as compared to that of $\sigma_1^{ts}$ ($\ket\psi_1^{ecs}$). Variation in depth of nonclassicality exposed by $\mathcal{D}_h^{(l-1)}$ is similar to the antibunching parameter. Also, $\mathcal{D}_h^{(l-1)}$ is not much affected by changing the values $q$ and $p$, that means a number of photons added or subtracted. Here for all values of $l$, the HOSPS parameter shows nonclassicality for both the states in a fixed range of $\bar{r}$ and $\alpha$. { Similar to the Mandel's $Q$ function and HOA, HOSPS also fails to detect the nonclassicality of original thermal or even coherent states.}

\subsection{Higher-order squeezing}
	
\begin{figure}[htb]
		\centering
		\includegraphics[scale=.9]{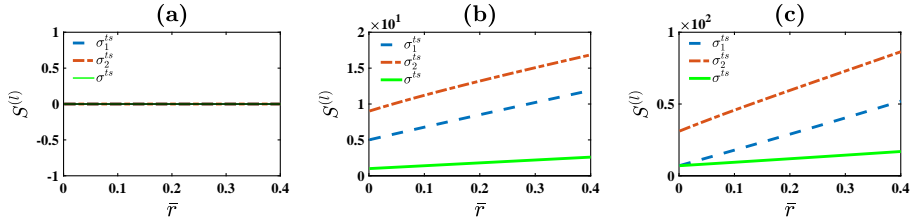}
		\caption{(Color online) $S^{(l)}$ with respect to mean photon number $\bar{r}$ for (a) $l=2,\,p=1,\,q=1$, (b) $l=4,\,p=1,\,q=2$ and (c) $l=6,\,p=2,\,q=1$.}
		\label{sltfig}
	\end{figure}

\begin{figure}[htb]
		\centering
		\includegraphics[scale=.9]{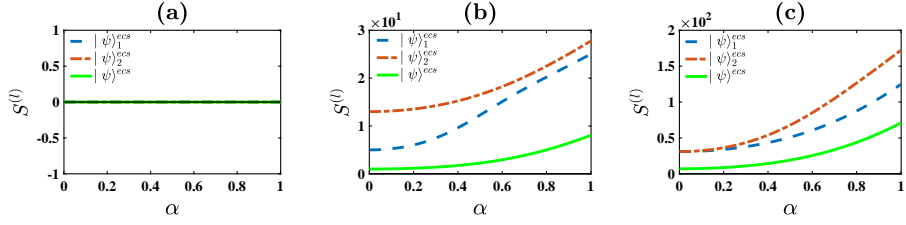}
		\caption{(Color online) $S^{(l)}$ with respect to coherent amplitude $\alpha$ for (a) $l=2,\,p=1,\,q=1$, (b) $l=4,\,p=1,\,q=2$ and (c) $l=6,\,p=2,\,q=1$.}
		\label{slecfig}
	\end{figure}

	The pioneering work of Hong and Mandel \cite{hong} introduced the HOS parameter to find out the nonclassicality of a given quantum state. HOS for even order \cite{priyam} is calculated when $l$-th order coherent state value is more than the field quadrature operator's moment. Thus HOS can be obtained by the inequality
	
	\begin{equation}
		\label{sl}
		S^{(l)} = \frac{\langle (\Delta X)^l \rangle -{\left(\frac{1}{2}\right)}_{\left(\frac{l}{2}\right)}}{{\left(\frac{1}{2}\right)}_{\left(\frac{l}{2}\right)}}\,\,<\,\,0,
	\end{equation}
	or
	
	\begin{equation}
		\label{sl1}
		\langle (\Delta X)^l \rangle\,\,<\,\,{\left(\frac{1}{2}\right)}_{\left(\frac{l}{2}\right)} = \frac{1}{2^{\frac{l}{2}}}(l-1)!!,
	\end{equation}
	with
	\begin{eqnarray}
		\label{sl2}
		\langle (\Delta X)^l \rangle = \sum_{r=0}^{l}\sum_{i=0}^{\frac{r}{2}}\sum_{k=0}^{r-2i}(-1)^r\frac{1}{2^\frac{1}{2}}(2i-1)!^{2i}C_k^lC_r^rC_{2i}\langle a^\dagger +a\rangle^{l-r}\langle a^{\dagger k}a^{r-2i-k}\rangle,
	\end{eqnarray}
	where $l$ is an even number and
	\begin{eqnarray*}
		n!!=
		\left\{
		\begin{array}{lll}
			& n(n-2)(n-4)\ldots 4.2 & \mbox{if $n$ is even},\\\\
			& n(n-2)(n-4)\ldots3.1 & \mbox{if $n$ is odd},
		\end{array}
		\right.
	\end{eqnarray*}
	The analytic expressions for the Hong-Mandel type HOS can be obtained by substituting \eqref{exppast}, \eqref{exppsat}, \eqref{eq:exppasec} and \eqref{eq:exppsaec} in \eqref{sl}-\eqref{sl2} for the states $\sigma_1^{ts}$, $\sigma_2^{ts}$, $\ket\psi_1^{ecs}$ and $\ket\psi_2^{ecs}$, respectively and the simplified results are plotted in Figs. \ref{sltfig} and \ref{slecfig}.
	
	
From the Figs.~\ref{sltfig} and \ref{slecfig}, it is clear that the HOS fails to display nonclassicality for any of the states  $\sigma_1^{ts}$, $\sigma_2^{ts}$, $\ket\psi_1^{ecs}$ and $\ket\psi_2^{ecs}$, as there is no negative value of $S^{(l)}$ in any of the parametric region of $\alpha$ as well as $\bar{r}$ \cite{priyam,ppr1,ppr2}. { HOS cannot detect the nonclassicality of original thermal or even coherent states in either case.}
	
	\subsection{Husimi-$Q$ function}
	
	A direct phase-space description of a quantum mechanical system is not possible due to the uncertainty principle. This leads to the creation of quasiprobability distributions, which are extremely helpful in quantum mechanics because they give a quantum-classical connection and make it easier to calculate quantum mechanical averages in a manner similar to classical phase-space averages \cite{kishore3}. The $Q$ function is one such quasiprobability distribution, and its zeros indicate nonclassicality \cite{ho48}. The Husimi-$Q$ function is defined as
	\begin{equation}\nonumber
		Q = \frac{1}{\pi}\bra{\beta}\sigma\ket{\beta},
	\end{equation}
	where $\ket{\beta}$ is the usual coherent state. A simple calculation finds the Husimi-$Q$ for $\sigma_1^{ts},\,\sigma_2^{ts},\, \ket\psi_1^{ecs}$ and $\ket\psi_2^{ecs}$ as following (see \ref{a2} for detailed calculations):
	
	\begin{eqnarray}
		Q_1^{ts} =\left\{
		\begin{array}{lll}
			\frac{N_1^{ts}q!^2e^{-|\beta|^2}|\beta|^{2(q-p)}}{\pi(1+\bar{r})(q-p)!^2}~_2F_2\left(1+q,1+q;1+q-p,1+q-p;\frac{\bar{r}|\beta|^{2}}{1+\bar{r}}\right)
			
			& \mbox{if} & q\geq p \\\\
			\frac{N_1^{ts}p!^2e^{-|\beta|^2}}{\pi(1+\bar{r})(p-q)!}~_2F_2\left(1+p,1+p;1,p-q+1;\frac{\bar{r}|\beta|^{2}}{1+\bar{r}}\right)
			
			& \mbox{if} & p >q
		\end{array}
		\right.
		\label{qpast}
	\end{eqnarray}

	\begin{eqnarray}
		Q_2^{ts} =\frac{N_2^{ts}e^{-|\beta|^2}|\beta|^{2(q-p)}}{\pi(1+\bar{r})}\left(\frac{\bar{r}|\beta|^2}{1+\bar{r}}\right)^pp!~_1F_1\left(1+p;1;\frac{\bar{r}|\beta|^2}{1+\bar{r}}\right)
		\label{qpsat}
	\end{eqnarray}

	\begin{eqnarray}
		Q_1^{ecs} =\frac{(N_1^{ecs})^2}{\pi}\exp(-|\alpha|^2-|\beta|^2)\left|H_{p,q}(\alpha,-\beta^*)(\exp(\alpha\beta^*)+(-1)^{p-r}\exp(-\alpha\beta^*))\right|^2
		\label{qpasec}
	\end{eqnarray}

	\begin{eqnarray}
		Q_2^{ecs} =\frac{(N_2^{ecs})^2}{\pi}|\beta|^{2q}|\alpha|^{2p}\exp(-|\alpha|^2-|\beta|^2)\left|\exp(\alpha\beta^*)+(-1)^p\exp(-\alpha\beta^*)\right|^2
		\label{qpsaec}
	\end{eqnarray}

\begin{figure}[htb]
		\centering
		\includegraphics[scale=0.9]{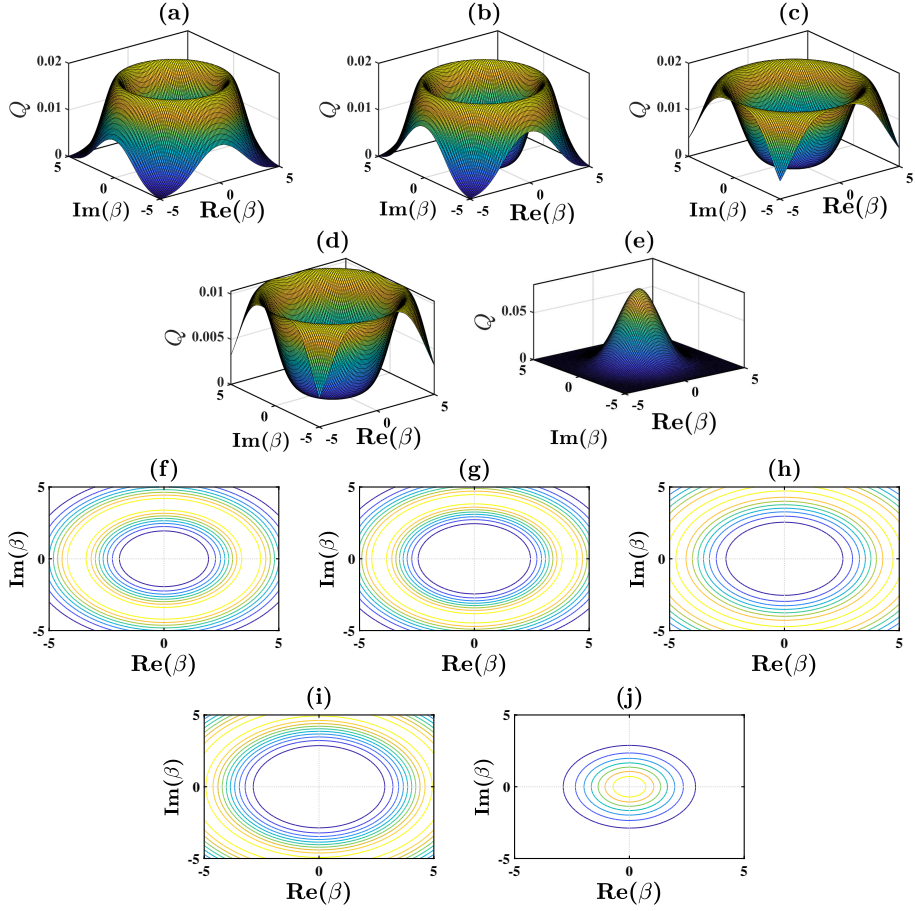}
		\caption{(Color online) Surface plots of $Q$ with respect to Re($\beta$) and Im($\beta$) for $p=2,\,q=4,\,\bar{r}=2$, (a) $\sigma_1^{ts}$, (b) $\sigma_2^{ts}$ and $p=4,\,q=2,\,\bar{r}=4$ and (c) $\sigma_1^{ts}$, (d) $\sigma_2^{ts}$ and $p=q=0$ and (e) $\sigma^{ts}$ along with corresponding contour plots in (f)-(j).}
	\label{qtfig}
\end{figure}

\begin{figure}[htb]
	\centering
	\includegraphics[scale=0.9]{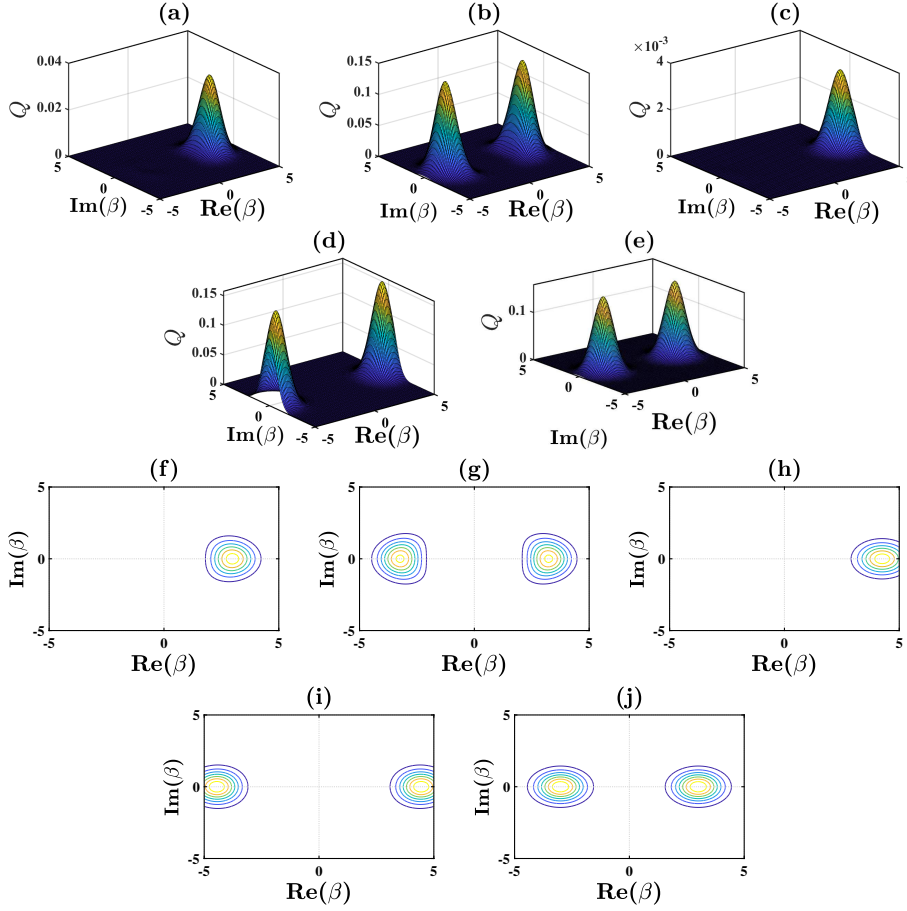}
	\caption{(Color online) Surface plots of $Q$ with respect to Re($\beta$) and Im($\beta$) for $p=2,\,q=4,\,\alpha=2$ and (a) $\ket\psi_1^{ecs}$, (b) $\ket\psi_2^{ecs}$, $p=4,\,q=2,\,\alpha=4$ and (c) $\ket\psi_1^{ecs}$, (d) $\ket\psi_2^{ecs}$, $p=q=0$ and (e) $\ket{\psi}^{ecs}$ along with corresponding contour plots in (f)-(j).}
\label{qecfig}
\end{figure}

	The zeroes of Husimi-$Q$ are obtained from  \eqref{qpast}-\eqref{qpsaec} for $\sigma_1^{ts}$, $\sigma_2^{ts}$, $\ket\psi_1^{ecs}$ and $\ket\psi_2^{ecs}$ and plotted in Figs. \ref{qtfig} and \ref{qecfig}.

It can be observed from Figs.~\ref{qtfig} and \ref{qecfig} that $\sigma_1^{ts}$, $\sigma_2^{ts}$, $\ket\psi_1^{ecs}$ and $\ket\psi_2^{ecs}$ indicate nonclassicality as there are  many points on which $Q=0$. The state $\sigma_2^{ts}$ ($\ket\psi_1^{ecs}$) has more points meeting the zero level as compared to $\sigma_1^{ts}$ ($\ket\psi_2^{ecs}$), that means $\sigma_2^{ts}$ ($\ket\psi_1^{ecs}$) state is more nonclassical as compared to $\sigma_1^{ts}$ ($\ket\psi_2^{ecs}$). With the increasing (decreasing) values of the state parameter $\bar{r}$ ($\alpha)$ and $p$ and decreasing (increasing) values of $q$, the number of zeros of $Q$ increases (decreases) and hence nonclassicality increases (decreases). { The zeros of the Husimi-$Q$ function depict the nonclassicality of original thermal and even coherent states but the depth of nonclassicality is less as compared to the PAS and PSA-operated states.}

\subsection{Agarwal-Tara criterion}

Agarwal and Tara defined a moment-based parameter for detecting the nonclassicality of an arbitrary quantum state, as $m_i=\langle a^{\dagger i}a^i\rangle$, $\forall\, i$. The zero-th order moment is one and the $k$-{th} order number distribution is $\mu_k = m_i^k$. Using these two, we derive the analytical expression for $A_3$ as
\begin{equation}
	\label{atc}
	A_3 = \frac{{\mbox{det}\,\,m}^{(3)}}{{\mbox{det}\,\,\mu}^{(3)} - {\mbox{det}\,\,m}^{(3)}}\,\, <\,\, 0,
\end{equation}
where $m_{i,j}^{(3)} = m_{i+j-2}$ with $m_0=1,\,m_i=\braket{a^{\dagger i}a^i}$ and  $\mu_{i,j}^{(3)} = \mu_{i+j-2}=m_1^{i+j-2}$.\\ In matrix form\\
\begin{equation*}
m^{(3)}=\left[
\begin{matrix}
1&m_1&m_2\\m_1&m_2&m_3\\m_2&m_3&m_4
\end{matrix}
\right]
\end{equation*}
and
\begin{equation*}
\mu^{(3)}=\left[
\begin{matrix}
1&\mu_1&\mu_2\\\mu_1&\mu_2&\mu_3\\\mu_2&\mu_3&\mu_4
\end{matrix}
\right]
\end{equation*}

\begin{figure}[htb]
	\centering
	\includegraphics[scale=.9]{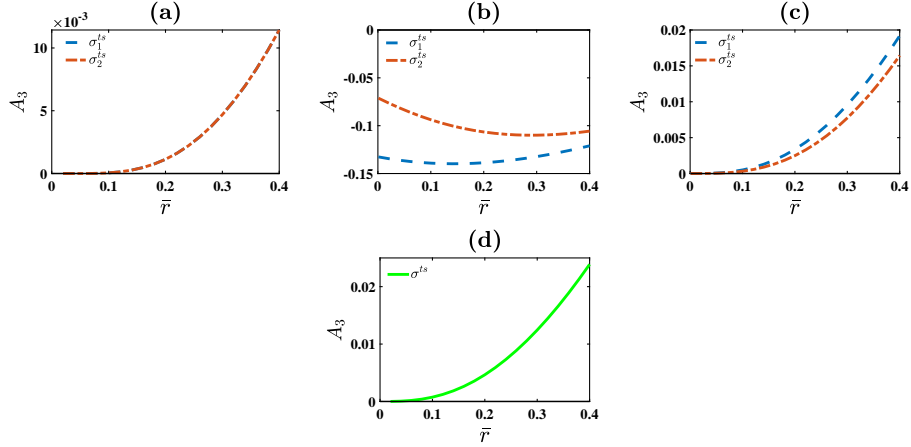}
	\caption{(Color online) $A_3$ with respect to $\bar{r}$ for (a) $p=1,\,q=1$, (b) $p=1,\,q=2$ (c) $p=2,\,q=1$ and (d) $p=q=0$.}
	\label{atctfig}
\end{figure}

\begin{figure}[htb]
	\centering
	\includegraphics[scale=.9]{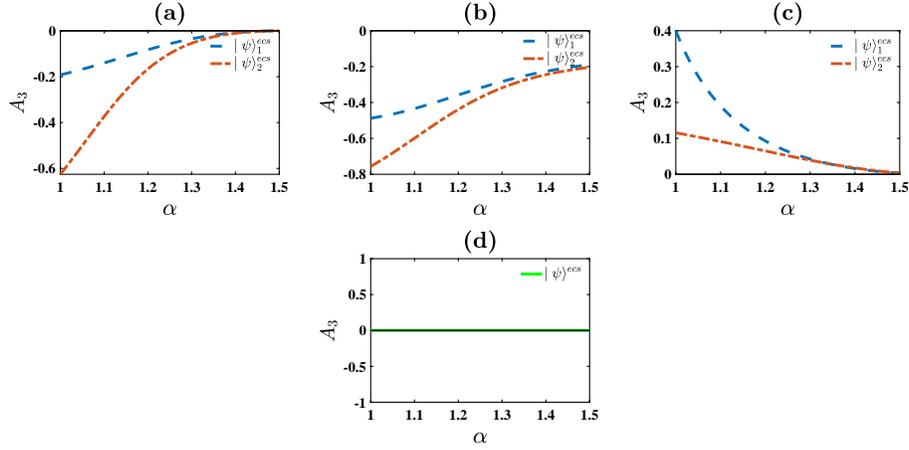}
	\caption{(Color online) $A_3$ with respect to $\alpha$ for (a) $p=1,\,q=1$, (b) $p=1,\,q=2$ (c) $p=2,\,q=1$ and (d) $p=q=0$.}
	\label{atcecfig}
\end{figure}

The expression in \eqref{atc} is simplified with the help of \eqref{exppast}, \eqref{exppsat}, \eqref{eq:exppasec} and \eqref{eq:exppsaec} for $\sigma_1^{ts}$, $\sigma_2^{ts}$, $\ket\psi_1^{ecs}$  and $\ket\psi_2^{ecs}$ and plotted in Figs.~\ref{atctfig} and \ref{atcecfig}.

From these figures, it can be found that $\sigma_1^{ts}$, $\sigma_2^{ts}$, $\ket\psi_1^{ecs}$ and $\ket\psi_2^{ecs}$ are nonclassical. When the number of photons added and subtracted is the same, then both $\sigma_1^{ts}$ and $\sigma_2^{ts}$ behave exactly similar way (see Fig. \ref{atctfig}(a)) but when the number of photons added is more than the number of photons subtracted then $\sigma_1^{ts}$ is more nonclassical than $\sigma_2^{ts}$ due to more negative values of $A_3$ (see Fig.~\ref{atctfig}(b)). In case of number of photons added is less than the number of photons subtracted then $A_3$ fails to detect the nonclassicality of both PAST and PSAT states (see Fig.~ \ref{atctfig}(c)). It is seen from Fig.~\ref{atcecfig} that $\ket{\psi}_2^{ecs}$ is more nonclassical than $\ket{\psi}_1^{ecs}$ for all values of $\alpha$ when the number of photons added is equal to or greater than the number of photons subtracted. Further the depth of nonclassicality for PSAEC is more than PASEC and it decreases with an increase in $\alpha$. { Clearly, $A_3$ fails to observe the nonclassicality of both original thermal and even coherent states.}

\subsection{Klyshko's criterion}

\begin{figure}[htb]
	\centering
	\includegraphics[scale=.9]{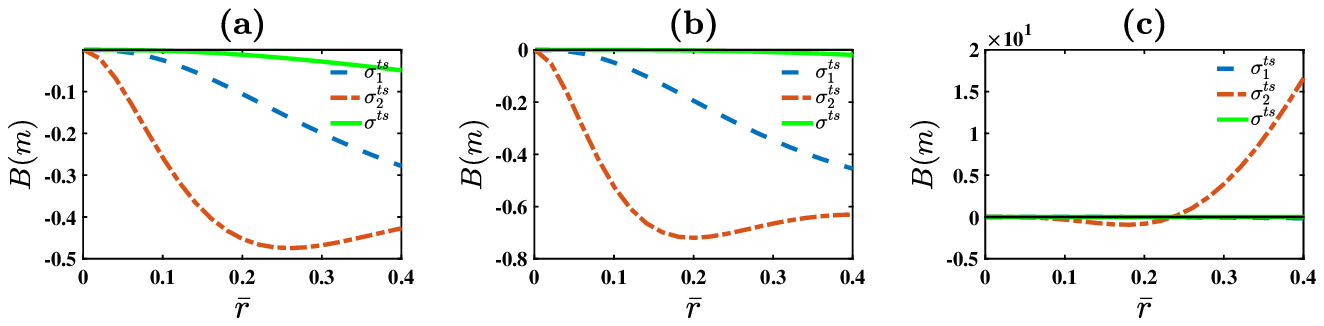}
	\caption{(Color online) $B(m)$ with respect to $\bar{r}$ for (a) $m=2,\,p=1,\,q=1$, (b) $m=3,\,p=1,\,q=2$ and (c) $m=4,\,p=2,\,q=1$.}
	\label{bmtfig}
\end{figure}

\begin{figure}[htb]
	\centering
	\includegraphics[scale=.9]{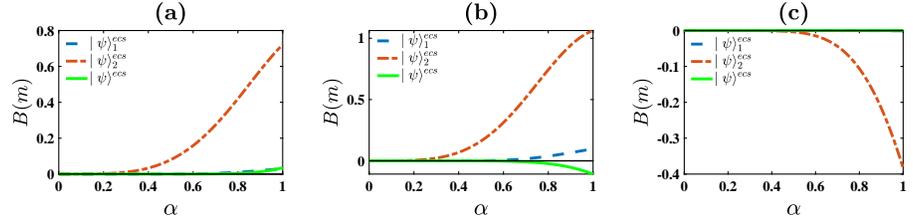}
	\caption{(Color online) $B(m)$ with respect to $\alpha$ for (a) $m=2,\,p=1,\,q=1$, (b) $m=3,\,p=1,\,q=2$ and (c) $m=4,\,p=2,\,q=1$.}
	\label{bmecfig}
\end{figure}

One of the most important criteria to detect the nonclassicality is given by Klyshko \cite{ho66}, who used three consecutive photon-number probabilities defined as
\begin{equation}
	\label{pm}
	p_m = \langle{{m}|\sigma|{m}}\rangle,
\end{equation}
$\sigma$ is the density matrix of the state under investigation.
Klyshko's inequality is
\begin{equation}
	\label{bm}
	B(m) = (m+2)p_mp_{m+2} - (m+1)(p_{m+1})^2\,\, <\,\,0
\end{equation}
Using \eqref{exppast}, \eqref{exppsat}, \eqref{eq:exppasec}, \eqref{eq:exppsaec} and \eqref{pm}, the functions $p_m$ for  the states are obtained as follows:

\begin{equation}
	p_{m_1}^{ts} = \frac{N_1^{ts}}{1+\bar{r}}\left(\frac{\bar{r}}{1+\bar{r}}\right)^{m+p-q}\frac{(m+p)!^2}{(m+p-q)!m!}
	\label{pmpast}
\end{equation}
\begin{equation}
	p_{m_2}^{ts} = \frac{N_2^{ts}}{1+\bar{r}}\left(\frac{\bar{r}}{1+\bar{r}}\right)^{m-q+p}\frac{(m-q+p)!m!}{(m-q)!}
	\label{pmpsat}
\end{equation}
\begin{equation}
	p_{m_1}^{ecs} =(N_1^{ecs})^2e^{-|\alpha|^2}|\alpha|^{2(m-q+p)}\left|\sum_{r=0}^{\text{min}(p,q)}{p\choose r}{q\choose r}r!\frac{\sqrt{m!}}{(m-q+r)!}(1+(-1)^{m-q+p})\right|^2
	\label{pmpasec}
\end{equation}
\begin{equation}
	p_{m_2}^{ecs} = (N_2^{ecs})^2|\alpha|^{2(m-q+p)}e^{-|\alpha|^2}\frac{m!}{(m-q)!^2}(1+(-1)^{m-q+p})^2
	\label{pmpsaec}
\end{equation}
Now using \eqref{bm}-\eqref{pmpsaec} we can find $B(m)$ and plot in the Figs.~\ref{bmtfig} and \ref{bmecfig}.


From the Figs.~\ref{bmtfig} and \ref{bmecfig}, it is clear that $\sigma_1^{ts}$, $\sigma_2^{ts}$, $\ket\psi_1^{ecs}$ and $\ket\psi_2^{ecs}$ states are nonclassical. Fig.~\ref{bmtfig} shows that $\sigma_2^{ts}$ is more nonclassical than $\sigma_1^{ts}$ for all values of $m$, $p$, $q$ but the nonclassicality decreases when photon addition is less than photon subtraction. This criterion fails to detect the nonclassicality of $\ket\psi_1^{ecs}$ and $\ket\psi_2^{ecs}$ states when photon addition is equal to or greater than photon subtraction (see Fig.~\ref{bmecfig}(a)-(b)). If the number of photons added is less than subtracted then $B(m)$ exposes the nonclassical character of PSAEC and PASEC states and also the depth of nonclassicality of PSAEC is greater than that of PASEC (see Fig.~\ref{bmecfig}(c)). { It is clear that only Klyshko's criteria detects the nonclassicality of thermal and even coherent states that decreases (increases) with $m$ for all thermal (even coherent) states.}

\section{Result}
\label{res}

It is clear that all the four states under consideration, namely $\sigma_1^{ts}$, $\sigma_2^{ts}$, $\ket\psi_1^{ecs}$ and $\ket\psi_2^{ecs}$ are nonclassical. The Mandel's criteria detects the nonclassicality of PSA-operated states but for some other states, this criteria fails to detect the nonclassical region, specially when $p>q$. HOA and HOSPS criteria locate the higher-order nonclassicality of PSA-operated states for all values of $l$, $p$, and $q$. These two criteria also detect that PSA-operated states are more highly nonclassical as compared to PAS-operated states. Further, the HOS criterion fails to detect the nonclassicality of any of the four states $\sigma_1^{ts}$, $\sigma_2^{ts}$, $\ket\psi_1^{ecs}$ and $\ket\psi_2^{ecs}$, as there is nowhere negativity of $S^{(l)}$ for all values of $l,\,p,\,q$ and state parameters $\bar{r}$ and $\alpha$. The Husimi-$Q$ function identifies the nonclassicality of the considered states but according to this criterion, the PSA-operated thermal (even coherent) state is more (less) nonclassical than the PAS-operated thermal (even coherent) state. This means the PSA operator results in more nonclassicality if the initial state is a thermal state and in the case of an even coherent state, the PAS operator causes more nonclassicality.
Agarwal-Tara condition behaves like Mandel's $Q_M$ and fails to detect nonclassicality of $\sigma_1^{ts}$ and $\sigma_2^{ts}$ for the case $p=q$. Klyshko's criterion reveals the nonclassical nature of $\sigma_1^{ts}$ and $\sigma_2^{ts}$ for all values of $m,\,p,\,q$ and $\bar{r}$ but nonclassicality of of $\ket\psi_1^{ecs}$ and $\ket\psi_2^{ecs}$ can be detected only for higher values of $m$ and $p>q$. In all the cases these criteria detect that PSA-operated states are more nonclassical than PAS-operated states. It can be concluded that PAS-operated states are less nonclassical than PSA-operated states except in the case of the Husimi-$Q$ function. In the previous section, where we discuss all the cases of nonclassicality, we observe that in general, all higher-order nonclassicality criteria show that PAST is more nonclassical than PSAT except in a few cases like PSAT is more nonclassical as compared to PAST detected by HOA for $l=3$, HOSPS for $l=3$, $S^{(l)}$ which does not show nonclassicality etc. Similarly, we have some criteria which show PSAEC is more nonclassical than PASEC such as Mandel's $Q$-function for $l=3$, $S^{(l)}$ which does not show non-classicality, Agarwal-Tara condition for $p=2,\,q=4$ and Klyshko's criteria for $m=4,\,6$ etc. All other higher-order criteria show PASEC is more nonclassical than PSAEC. Thus by comparing both the case of applying PAS and PSA on thermal and even coherent states, we can easily conclude that PAS-operator gives more nonclassical state than PSA-operator. Hence if we first add and then subtract photons from an arbitrary state most likely in thermal and even coherent states, it will produce more nonclassicality than first subtracting and then adding the same number of photons to the quantum state. { In addition, the nonclassicality of thermal and even coherent states is detected only by the Husimi-$Q$ function and Klyshko's criteria.}

\begin{appendix}
	\section{Detailed calculation for $\langle a^{\dagger m}a^n\rangle$ with respect to different states}
	
	\label{a1}
	\begin{align*}
		E_1^{ts}&=
		Tr(a^{\dagger m}a^n\sigma_1^{ts})\\
&=Tr\left( \frac{N_1^{ts}}{1+\bar{r}}\sum_{r=0}^{\infty}\left(\frac{\bar{r}}{1+\bar{r}}\right)^r\frac{(r+q)!^2}{r!(r+q-p)!}a^{\dagger m}a^n\ket{r+q-p}\bra{r+q-p}\right)
		\\&=Tr\left( \frac{N_1^{ts}}{1+\bar{r}}\sum_{r=0}^{\infty}\left(\frac{\bar{r}}{1+\bar{r}}\right)^r\frac{(r+q)!^2}{r!(r+q-p)!}\sqrt{\frac{(r+q-p)!}{(r+q-p-n)!}}\times\right.
\\&\left.\sqrt{\frac{(r+q-p-n+m)!}{(r+q-p-n)!}}\ket{r+q-p-n+m}\bra{r+q-p}\right)
\\&= \frac{N_1^{ts}}{1+\bar{r}}\sum_{r=0}^{\infty}\left(\frac{\bar{r}}{1+\bar{r}}\right)^r\frac{(r+q)!^2}{r!(r+q-p-n)!}\delta_{m,n}\\&=\left\{
		\begin{array}{ll}
			\frac{N_1^{ts}}{1+\bar{r}}\sum_{r=0}^{\infty}\left(\frac{\bar{r}}{1+\bar{r}}\right)^r\frac{(r+q)!^2}{r!(r+q-p-n)!}\delta_{m,n} & \mbox{if } q-p-n\geq 0\\
			\frac{N_1^{ts}}{1+\bar{r}}\sum_{r=p-q+n}^{\infty}\left(\frac{\bar{r}}{1+\bar{r}}\right)^r\frac{(r+q)!^2}{r!(r+q-p-n)!}\delta_{m,n} & \mbox{if } q-p-n < 0
		\end{array}
		\right. \\&=\left\{
		\begin{array}{ll}
			\frac{N_1^{ts}}{1+\bar{r}}\sum_{r=0}^{\infty}\left(\frac{\bar{r}}{1+\bar{r}}\right)^r\frac{(r+q)!^2}{r!(r+q-p-n)!}\delta_{m,n} & \mbox{if } q-p-n\geq 0\\
			\frac{N_1^{ts}}{1+\bar{r}}\sum_{r=0}^{\infty}\left(\frac{\bar{r}}{1+\bar{r}}\right)^{r-q+p+n}\frac{(r+p+n)!^2}{r!(r-q+p+n)!}\delta_{m,n} & \mbox{if } q-p-n < 0
		\end{array}
		\right. \\&=\left\{
		\begin{array}{lll}
			\frac{N_1^{ts}q!^2}{(q-p-n)!(1+\bar{r})}\\\times~_2F_1\left(1+q,1+q;q-p-n+1;\frac{\bar{r}}{1+\bar{r}}\right)
			\delta_{m,n} & \mbox{if } & q-p-n\geq 0\\
			\frac{N_1^{ts}(p+n)!^2}{(p-q+n)!(1+\bar{r})}\left(\frac{\bar{r}}{1+\bar{r}}\right)^{p-q+n}\\\times~_2F_1\left(1+p+n,1+p+n;p-q+n+1;\frac{\bar{r}}{1+\bar{r}}\right)
			\delta_{m,n} & \mbox{if } & q-p-n < 0,
		\end{array}
		\right.
	\end{align*}
	where $ _pF_q(a_i;b_j,c)$ is the well-known generalized hypergeometric function.
	\begin{align*}
		E_2^{ts}&
		=\frac{N_2^{ts}}{1+\bar{r}}\sum_{r=0}^{\infty}\left(\frac{\bar{r}}{1+\bar{r}}\right)^r\frac{r!(r-p+q)!^2}{(r-p)!^2(r-p+q-n)!}\delta_{m,n} \\&=\left\{
		\begin{array}{ll}
			\frac{N_2^{ts}}{1+\bar{r}}\sum_{r=p}^{\infty}\left(\frac{\bar{r}}{1+\bar{r}}\right)^r\frac{r!(r-p+q)!^2}{(r-p)!^2(r-p+q-n)!}\delta_{m,n} &\mbox{if } q\geq n \\
			\frac{N_2^{ts}}{1+\bar{r}}\sum_{r=p-q+n}^{\infty}\left(\frac{\bar{r}}{1+\bar{r}}\right)^r\frac{r!(r-p+q)!^2}{(r-p)!^2(r-p+q-n)!}\delta_{m,n} &\mbox{if } q < n
		\end{array}
		\right.
		\\&=\left\{
		\begin{array}{lll}
			\frac{N_2^{ts}}{1+\bar{r}}\left(\frac{\bar{r}}{1+\bar{r}}\right)^p\frac{q!^2p!}{(q-n)!}~_3F_2\left(1+q,1+q,1+p;1,q-n+1;\frac{\bar{r}}{1+\bar{r}}\right)
			\delta_{m,n} &\mbox{if } q\geq n \\
			\frac{N_2^{ts}}{1+\bar{r}}\left(\frac{\bar{r}}{1+\bar{r}}\right)^{p-q+n}\frac{n!^2(p-q+n)!}{(n-q)!^2}\times\\_3F_2\left(1+n,1+n,1+p-q+n;1-q+n,1-q+n;\frac{\bar{r}}{1+\bar{r}}\right)
			\delta_{m,n} &\mbox{if } q < n
		\end{array}
		\right.
	\end{align*}
	Now first we write the expansion $a^pa^{\dagger m}a^na^{\dagger p}$ in normal order as follows
	\begin{align*}
		a^pa^{\dagger m}a^na^{\dagger p}&=a^pa^{\dagger m}\sum_{r=0}^{\text{min}(n,p)}{n\choose r}{p\choose r}r!a^{\dagger p-r}a^{n-r}\\&=\sum_{r=0}^{\text{min}(n,p)}{n\choose r}{p\choose r}r!a^pa^{\dagger m+p-r}a^{n-r}\\&= \sum_{r=0}^{\text{min}(n,p)}{n\choose r}{p\choose r}r!\sum_{s=0}^{\text{min}(p,m+p-r)}{p\choose s}{m+p-r \choose s}s!a^{\dagger m+p-r-s}a^{p-s}a^{n-r}\\&=\sum_{r=0}^{\text{min}(n,p)}{n\choose r}{p\choose r}r!\sum_{s=0}^{\text{min}(p,m+p-r)}{p\choose s}{m+p-r \choose s}s!a^{\dagger m+p-r-s}a^{n+p-r-s}
	\end{align*}
	\begin{align*}
		\langle a^{\dagger m}a^n\rangle_1^{ecs}&=\bra{\psi}_1^{ecs}a^{\dagger m}a^n\ket{\psi}_1^{ecs}
		\\&= (N_1^{ecs})^2\bra\psi a^qa^{\dagger p+m}a^{n+p}a^{\dagger q}\ket\psi 
		\\&=(N_1^{ecs})^2\sum_{r=0}^{\text{min}(n+p,q)}{n+p\choose r}{q\choose r}r!\sum_{s=0}^{\text{min}(q,m+p+q-r)}{q\choose s}{m+p+q-r \choose s}s!\\&\times({1+(-1)^{m+n}})\alpha^{*m+p+q-r-s}\alpha^{n+p+q-r-s}(1+(-1)^{m+p+q-r-s}e^{-2|\alpha|^2})
		\\&= (N_1^{ecs})^2\sum_{r=0}^{\text{min}(n+p,q)}{n+p\choose r}{q\choose r}r!\alpha^{n+p-r}(-1)^q\left(H_{m+p+q-r,q}(\alpha^*,-\alpha)\right.\\&\left.+e^{-2|\alpha|^2}H_{m+p+q-r,q}(-\alpha^*,-\alpha)\right)
	\end{align*}
	Similarly, we get
	\begin{align*}
		\langle a^{\dagger m}a^n\rangle_2^{ecs}&=
		(N_2^{ecs})^2\sum_{r=0}^{\text{min}(n,q)}{n\choose r}{q\choose r}r!\sum_{s=0}^{\text{min}(q,m+q-r)}{q\choose s}{m+q-r \choose s}s!\\&\times (1+(-1)^{m+n})\alpha^{*m+q+p-r-s}\alpha^{n+q+p-r-s}(1+(-1)^{m+q+p-r-s}e^{-2|\alpha|^2}) \\&=(N_2^{ecs})^2|\alpha|^{2q}\sum_{r=0}^{\text{min}(n,q)}{n\choose r}{q\choose r}r!\alpha^{n-r}(-1)^q\left(H_{m+q-r,q}(\alpha^*,-\alpha)\right.\\&\left.+e^{-2|\alpha|^2}H_{m+q-r,q}(-\alpha^*,-\alpha)\right)
	\end{align*}
	\section{Husimi-$Q$ function for considered states}
	\label{a2}
	\begin{align*}
		\pi Q_1^{ts}&= 
		\frac{N_1^{ts}}{1+\bar{r}}\sum_{r=0}^{\infty}\left(\frac{\bar{r}}{1+\bar{r}}\right)^r\frac{(r+q)!^2}{r!(r+q-p)!}\left|e^{-|\beta|^2/2|}\beta^{* r+q-p}/\sqrt{(r+q-p)!}\right|^2 
		\\&=\frac{N_1^{ts}e^{-|\beta|^2}|\beta|^{2(q-p)}}{1+\bar{r}}\sum_{r=0}^{\infty}\left(\frac{\bar{r}|\beta|^{2}}{1+\bar{r}}\right)^r\frac{(r+q)!^2}{r!(r+q-p)!^2} \\&=\left\{
		\begin{array}{ll}
			\frac{N_1^{ts}e^{-|\beta|^2}|\beta|^{2(q-p)}}{1+\bar{r}}\sum_{r=0}^{\infty}\left(\frac{\bar{r}|\beta|^{2}}{1+\bar{r}}\right)^r\frac{(r+q)!^2}{r!(r+q-p)!^2} & \mbox{if } q\geq p \\
			\frac{N_1^{ts}e^{-|\beta|^2}|\beta|^{2(q-p)}}{1+\bar{r}}\sum_{r=p-q}^{\infty}\left(\frac{\bar{r}|\beta|^{2}}{1+\bar{r}}\right)^r\frac{(r+q)!^2}{r!(r+q-p)!^2} & \mbox{if } p >q
		\end{array}
		\right. \\&=\left\{
		\begin{array}{ll}
			\frac{N_1^{ts}e^{-|\beta|^2}|\beta|^{2(q-p)}}{1+\bar{r}}\sum_{r=0}^{\infty}\left(\frac{\bar{r}|\beta|^{2}}{1+\bar{r}}\right)^r\frac{(r+q)!^2}{r!(r+q-p)!^2} & \mbox{if } q\geq p \\
			\frac{N_1^{ts}e^{-|\beta|^2}|\beta|^{2(q-p)}}{1+\bar{r}}\sum_{r=0}^{\infty}\left(\frac{\bar{r}|\beta|^{2}}{1+\bar{r}}\right)^{r-q+p}\frac{(r+p)!^2}{r!^2(r-q+p)!} & \mbox{if } p >q
		\end{array}
		\right. \\&=\left\{
		\begin{array}{ll}
			\frac{N_1^{ts}q!^2e^{-|\beta|^2}|\beta|^{2(q-p)}}{(1+\bar{r})(q-p)!^2}~_2F_2\left(1+q,1+q;1+q-p,1+q-p;\frac{\bar{r}|\beta|^{2}}{1+\bar{r}}\right)
			& \mbox{if } q\geq p \\
			\frac{N_1^{ts}p!^2e^{-|\beta|^2}}{(1+\bar{r})(p-q)!}~_2F_2\left(1+p,1+p;1,p-q+1;\frac{\bar{r}|\beta|^{2}}{1+\bar{r}}\right)
			& \mbox{if } p >q
		\end{array}
		\right.
	\end{align*}
	Similarly
	\begin{align*}
		\pi Q_2^{ts}&=\frac{N_2^{ts}}{1+\bar{r}}\sum_{r=0}^{\infty}\left(\frac{\bar{r}}{1+\bar{r}}\right)^r\frac{r!(r-p+q)!}{(r-p)!^2}e^{-|\beta|^2}|\beta|^{2(r-p+q)}/(r-p+q)! \\&=\frac{N_2^{ts}}{1+\bar{r}}\sum_{r=p}^{\infty}\left(\frac{\bar{r}}{1+\bar{r}}\right)^r\frac{r!(r-p+q)!}{(r-p)!^2}e^{-|\beta|^2}|\beta|^{2(r-p+q)}/(r-p+q)!\\&=\frac{N_2^{ts}e^{-|\beta|^2}|\beta|^{2(q-p)}}{1+\bar{r}}\sum_{r=0}^{\infty}\left(\frac{\bar{r}|\beta|^2}{1+\bar{r}}\right)^{r+p}\frac{(r+p)!}{r!^2}\\&=\frac{N_2^{ts}e^{-|\beta|^2}|\beta|^{2(q-p)}}{1+\bar{r}}\left(\frac{\bar{r}|\beta|^2}{1+\bar{r}}\right)^pp!~_1F_1\left(1+p;1;\frac{\bar{r}|\beta|^2}{1+\bar{r}}\right)
	\end{align*}
	\begin{align*}
		\pi Q_1^{ecs}&=
		(N_1^{ecs})^2\left|\bra{\beta}\sum_{r=0}^{\text{min}(p,q)}{p\choose r}{q\choose r}r!a^{\dagger q-r}a^{p-r}(\ket\alpha+\ket{-\alpha})\right|^2 
		\\&= (N_1^{ecs})^2\exp(-|\alpha|^2-|\beta|^2)\left|\sum_{r=0}^{\text{min}(p,q)}{p\choose r}{q\choose r}r!\beta^{* q-r}\alpha^{p-r}(\exp(\alpha\beta^*)+(-1)^{p-r}\exp(-\alpha\beta^*))\right|^2 
		\\&=(N_1^{ecs})^2\exp(-|\alpha|^2-|\beta|^2)\left|H_{p,q}(\alpha,-\beta^*)(\exp(\alpha\beta^*)+(-1)^{p-r}\exp(-\alpha\beta^*))\right|^2
	\end{align*}
	similarly $Q_2^{ecs}$ can be obtained.
\end{appendix}
\begin{center}
	\textbf{ACKNOWLEDGEMENT}
\end{center}
Deepak's work is supported by the Council of Scientific and Industrial Research (CSIR), Govt. of India (Award no. 09/1256(0006)/2019-EMR-1).

\end{document}